\documentclass[prx,aps,showpacs,twocolumn,preprintnumbers,
amsmath,amssymb,superscriptaddress,longbibliography,nofootinbib]{revtex4-1}
\usepackage[english]{babel}
\usepackage{amsmath,amssymb,amsfonts}
\usepackage{multirow,makecell,booktabs}
\usepackage{graphicx}
\usepackage[colorlinks=True,linkcolor=red,citecolor=blue,urlcolor=blue]{hyperref}
\usepackage{bookmark}
\usepackage[table,xcdraw]{xcolor}
\usepackage{physics}
\usepackage{nicefrac}
\usepackage{bm}
\usepackage{xspace}
\usepackage{bbm}
\usepackage{colortbl}
\usepackage{slashed}
\usepackage[normalem]{ulem}
\usepackage{tcolorbox}
\usepackage[caption=false]{subfig}
\usepackage{tikz}
\newcommand{\pytheus}{\textsc{pytheus}\xspace}
\newcommand{\halo}{\textsc{Halo}\xspace}
\newcommand{\mpl}{Max Planck Institute for the Science of Light, Erlangen, Germany}
\begin{document}
\title{Digital Discovery of a Scientific Concept at the Core of Experimental Quantum Optics}
\author{Sören Arlt}
 \email{soeren.arlt@mpl.mpg.de}
\affiliation{\mpl}
\author{Carlos Ruiz-Gonzalez}
\affiliation{\mpl}
\author{Mario Krenn}
 \email{mario.krenn@mpl.mpg.de}
\affiliation{\mpl}
\date{\today} 
\begin{abstract}
Entanglement is a crucial resource for quantum technologies ranging from quantum communication to quantum-enhanced measurements and computation. Finding experimental setups for these tasks is a conceptual challenge for human scientists due to the counterintuitive behavior of multiparticle interference and the enormously large combinatorial search space.
Recently, new possibilities have been opened by artificial discovery where artificial intelligence proposes experimental setups for the creation and manipulation of high-dimensional multi-particle entanglement.
While digitally discovered experiments go beyond what has been conceived by human experts, a crucial goal is to understand the underlying concepts which enable these new useful experimental blueprints.
Here, we present \textsc{Halo} (Hyperedge Assembly by Linear Optics), a new form of multiphoton quantum interference with surprising properties. \textsc{Halos} were used by our digital discovery framework to solve previously open questions. We -- the human part of this collaboration -- were then able to conceptualize the idea behind the computer discovery and describe them in terms of effective probabilistic multi-photon emitters. We then demonstrate its usefulness as a core of new experiments for highly entangled states, communication in quantum networks, and photonic quantum gates.
Our manuscript has two conclusions. First, we introduce and explain the physics of a new practically useful multi-photon interference phenomenon that can readily be realized in advanced setups such as integrated photonic circuits. Second, our manuscript demonstrates how artificial intelligence can act as a source of inspiration for the scientific discoveries of new actionable concepts in physics.
\end{abstract}
\maketitle
A central element of physics (and science in general) is the formulation of concepts. A range of related objects, processes or properties are unified under one name (such as \textit{atom}, \textit{scattering}, \textit{charge}), simplifying our description of the world. Using artificial intelligence (AI) to inspire and ultimately perform conceptualization could accelerate scientific progress greatly.
AI has been applied in physics to rediscover hidden symmetries \cite{hiddensymmetries}, orbital mechanics \cite{orbital} \cite{renner}, models for gravitational-wave populations \cite{gwave}, conserved quantities \cite{conserveredq} and phase diagrams \cite{phasediagrams}. These works show the great potential that lies in applying machine learning to the discovery of underlying concepts in physics. However, their results focus mostly on rediscovery and it is not per se clear whether and how to extend these results to the discovery of \textit{new} concepts and ideas.

Here we go beyond rediscovery, and use AI to discover a hidden concept that lies at the heart of quantum optics -- a field in which automated discovery and design have been strongly employed recently \cite{melvin, knott2016search, melnikov2018active, o2019hybrid, melnikov2020setting, theseus, cervera2021design, flam2022learning} (see a review in \cite{krenn2020computer}).
We obtain these results using \pytheus -- a highly efficient discovery framework that provides interpretable results, which is presented in a parallel article \cite{hundred} and is available as open-source software \footnote{\url{https://github.com/artificial-scientist-lab/Theseus}}. The concept, which we call \halo (Hyperedge Assembly by Linear Optics) is a multiphoton interference effect with a surprising property: It acts as a probabilistic source of multi-photon pairs, while being constructed from only pair sources under common experimental post-selection conditions. As we show, this resource can now be applied to construct by hand experimental setups for creating new forms of multi-particle entangled states, entanglement swapping, and quantum gates, which were previously beyond our intuition. Furthermore, we see that a rudimentary form of \textsc{Halos} can be found in several of the pioneering quantum information experiments with photons, which can now be understood as special cases of the much more general design principle \textsc{Halo}.

\begin{figure}[t]
    \includegraphics[width=0.48\textwidth]{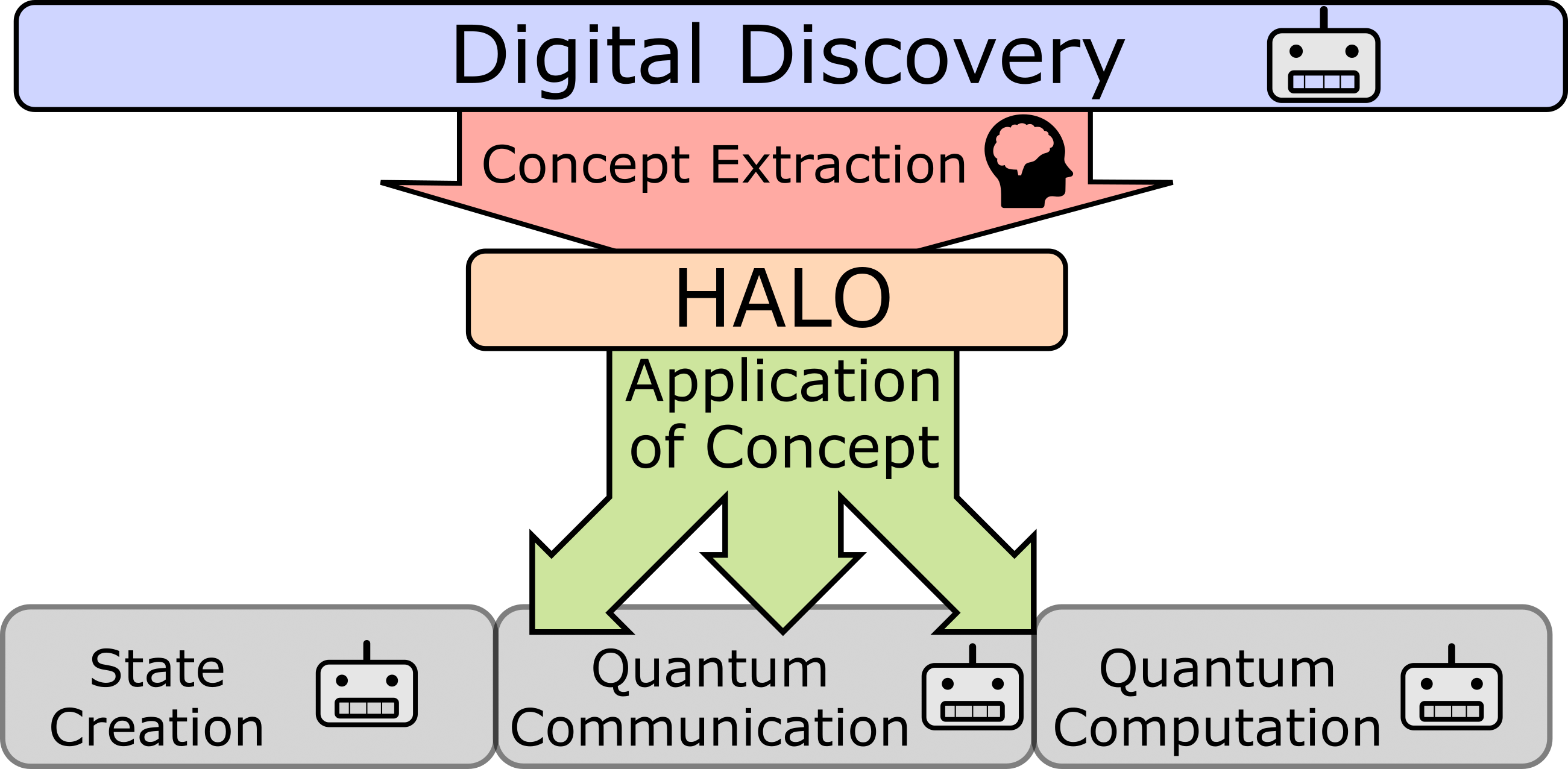}
    \caption{Structure of this article. The design principle \textsc{Halo} is formulated after analysis of the results found by \pytheus algorithm. The concept is applied in three different areas of quantum optics, where solutions discovered by the computer are extended by hand.}
    \label{fig:flowchart}
\end{figure}

At a broader view, we demonstrate how computer algorithms can act as a source of inspiration to increase our scientific understanding \cite{understanding}.
We illustrate the connections between the parts of this article in Fig. \ref{fig:flowchart}.

\begin{figure}
\centering
    \includegraphics[width=0.4\textwidth]{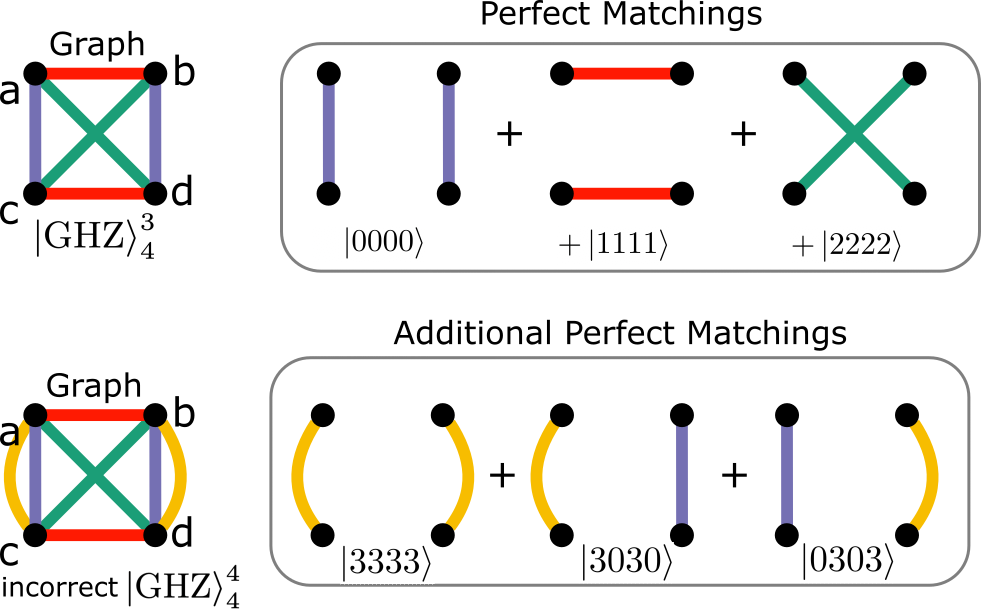}
    \caption{(a) A graph corresponding to the experimental setup for the creation of the $\ket{\text{GHZ}}_4^3$ state. A perfect matching is a set of edges by which each vertex of a graph is covered exactly once. The graph has three perfect matchings (blue, red, and green). In this case, each perfect matching corresponds to one term in the target state. (b) shows an incorrect attempt at drawing a graph corresponding to the state $\ket{\text{GHZ}}_4^4$. The two additional edges create the wanted fourth term at the cost of two unwanted cross-terms.}
    \label{fig:434}
\end{figure}

\textbf{Digital Discovery} -- 
The digital discovery framework \pytheus relies on a representation of quantum experiments using graphs \cite{graphs1}. This representation was first developed for entanglement by path identity \cite{pathidentity2,pathidentity} but also extends to bulk optics or integrated photonics \cite{graphs2,theseus}.  Integrated photonic chips have made impressive technological progress recently and offer great potential for the experimental realization of setups discovered by \pytheus \cite{photonicchips,integratedsilicon,natrev22integrated,Lu2021-to,Feng2020-wh,Zhang2019-gf}. The search for an experiment creating a target state is formulated as an optimization that maximizes the fidelity of a graph. When a solution is found, the corresponding graph can directly be trannslated to an experimetnal setup consisting of standard optical components. A detailed explanation of digital discovery using \pytheus is provided in the accompanying paper \cite{hundred}. Here we focus on the scientific consequences of the new quantum optics concept \textsc{Halo}, the underlying physics, its connection to modern quantum optics experiments, and how it can be productively used.\\
A prominent class of examples for genuine multi-particle entanglement are the GHZ states \cite{highdimghz1} \cite{highdimghz2}.
\begin{align}
    \ket{\text{GHZ}}_n^d = \frac{1}{\sqrt{d}}\sum_{i=0}^d \ket{i}^{\otimes n}
\end{align}
where $n$ is the number of particles and $d$ is the dimension of the GHZ state. $\ket{\text{GHZ}}_n^d$ is a generalization of the original $\ket{\text{GHZ}}_3^2$ state \cite{ghz}.
Fig. \ref{fig:434} shows a correlation network for the state $\ket{\text{GHZ}}_4^3$. An edge of the graph corresponds to a two-particle correlation as it is introduced by probabilistic photon pair sources, for example by spontaneous parametric down-conversion (SPDC). A vertex corresponds to a path to a detector at the end of the setup. We condition the final quantum state on the detection of one photon in each of the detectors, which motivates the usage of perfect matchings in graphs. A perfect matching is a set of edges by which each vertex of a graph is covered exactly once. In an event where all crystals corresponding to the edges in a perfect matching fire, exactly one photon will enter each detector. Fig. \ref{fig:434} also shows that each perfect matching of the graph can be understood as a contribution to the created state. \\ 
Fig. \ref{fig:434} also shows that it is not straightforward to extend the construction to $\ket{\text{GHZ}}_4^4$. It is impossible to create $\ket{\text{GHZ}}_4^4$ using linear optics without the use of additional resources in the form of ancillary photons \cite{graphs1,graphs2}. This is a physical limitation that can be explained by graph theory as follows. There are at most three disjoint perfect matchings in a four-vertex graph. This makes it impossible to create four different GHZ terms without introducing cross terms with a four-vertex graph.

\begin{figure*}[t]
    \includegraphics[width=\textwidth]{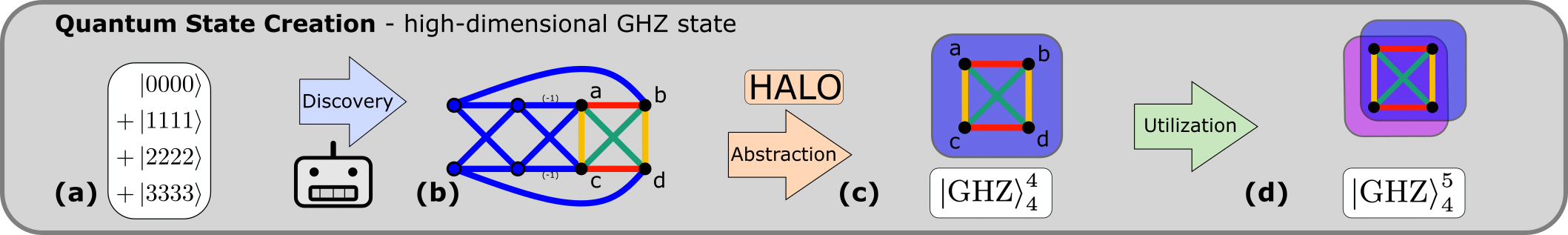}
    \caption{(a) shows the target state $\ket{\text{GHZ}}_4^4$ (b) shows the solution found by \pytheus for the $\ket{\text{GHZ}}_n^d$ state. The four additional vertices correspond to four ancillary particles necessary for this construction. (c) shows the solution for $\ket{\text{GHZ}}_4^4$ in the abstract \textsc{Halo} representation with the hyperedge imitated by the ancillary graph. (d) illustrates the utilization step. Knowing the ancillary graph that corresponds to the abstract representation with hyperedges this can be directly translated back to a graph and further into an experimental setup.}
    \label{fig:448hyper}
\end{figure*}

Including additional resources allows us to go beyond this limitation. We can use \pytheus to search for a graph corresponding to the state
\begin{align}
    \ket{\psi} = \ket{\text{GHZ}}_4^4 \otimes \ket{0000},
\end{align}
which is the four-dimensional four-particle GHZ state in a product state with four ancillary particles. For this target, \pytheus discovers minimal solution in Fig. \ref{fig:448hyper} (b). The solution shown has 12 perfect matchings. Four pairs interfere constructively creating the four GHZ terms. The remaining two pairs each correspond to cross-terms but interfere destructively.

\begin{figure}[t]
    \includegraphics[width=0.48\textwidth]{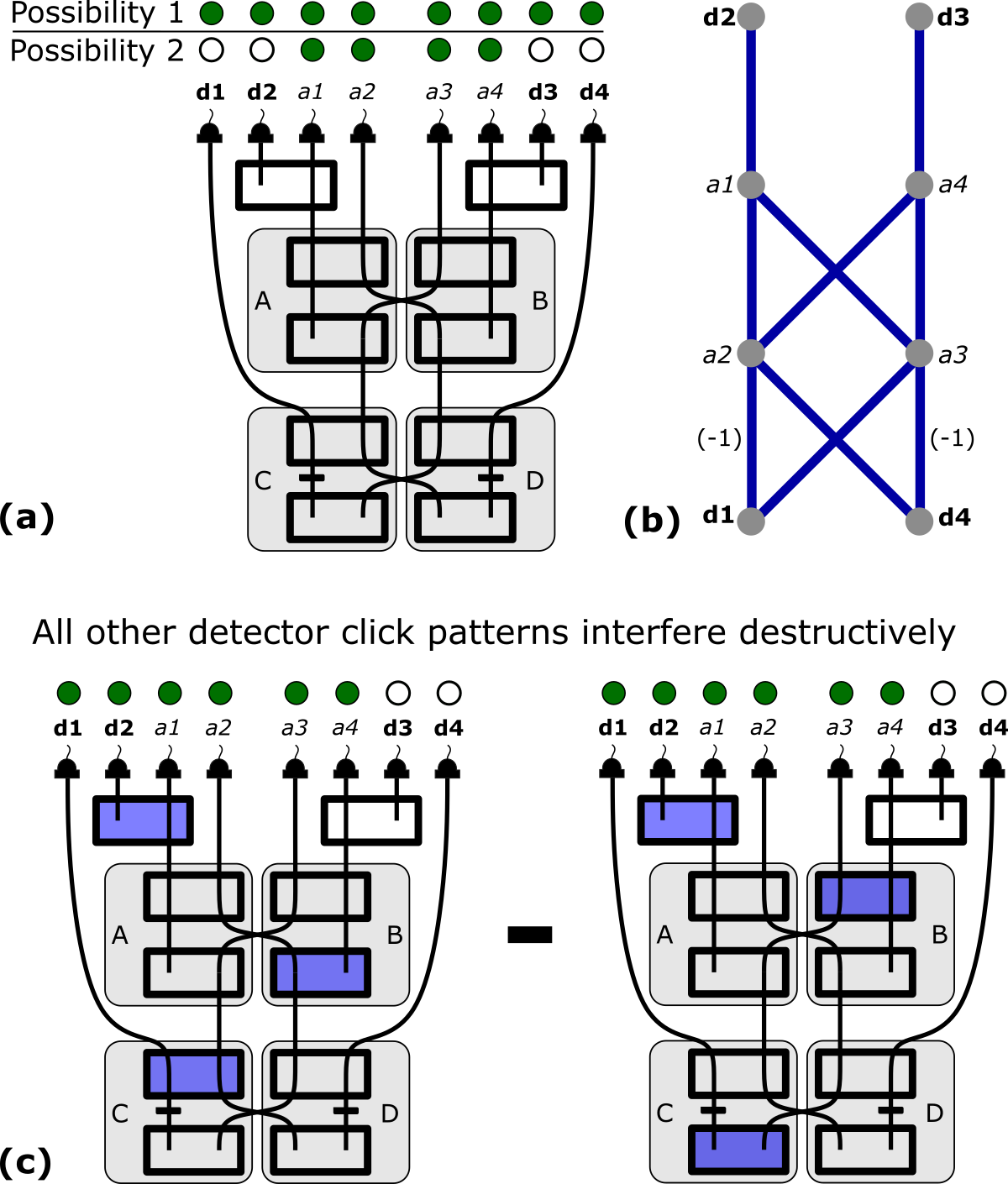}
    \caption{(a) Physical setup (entanglement by path identity) corresponding to (b) the ancillary graph for the state creation of $\ket{\text{GHZ}}_4^4$. The four-particle emitter is emulated in detectors \textbf{d1}-\textbf{d4}. Events with click pattern CP 1 interfere constructively (sub setup made from containers C and D and the two SPDC crystals at the top). Further, events with click pattern CP2 interfere constructively (sub setup made from containers A and B). Events that do not produce CP1 or CP2 cancel out through frustrated interference. This can be seen from the fact that the sub setup made from containers A and D as well as the sub setup made from containers B and C create frustrated interference. (c) shows an example of two events destructively interfering.}
    \label{fig:halophysical}
\end{figure}

\textbf{The physics of \textsc{H\footnotesize{ALO}}} -- By closer inspection of the solution for $\ket{\text{GHZ}}_4^4$ (Fig. \ref{fig:448hyper} (b)), we see that the graph for $\ket{\text{GHZ}}_4^3$ (shown in Fig. \ref{fig:434} (a)) is included as a subgraph. The experimental setup for $\ket{\text{GHZ}}_4^3$ can be seen as a basic setup that is extended by the components corresponding to the remaining edges of the graph $\ket{\text{GHZ}}_4^4$. When all detectors click, the additional components either produce four correlated particles or none, with all cross-terms destructively interfering, imitating a four-particle emitter. The physical interpretation of the ancillary subgraph is shown in Fig. \ref{fig:halophysical}. This is achieved by an interference pattern that can be interpreted as an extension of frustrated multiphoton interference, an effect described in \cite{graphs2} and experimentally observed in \cite{FIintegrated,FInonlocal}). Building probabilistic multi-particle emitters is an active area of research \cite{multiphoton,photongun}. \halo offers a way of emulating correlated multi-particle emitters with pair sources in post-selected experiments, which is a complementary experiment route that can employ physically well-understood technologies.

\textbf{Concept Extraction} -- We give the concept that appears in the solution for $\ket{\text{GHZ}}_4^4$ a name:

\textit{A \textsc{Halo} (Hyperedge Assembly by Linear Optics) is a subsystem of a linear optics setup, which effectively acts as a probabilistic multi-photon source.}

This definition is not used by the algorithm to produce solutions, rather we abstract it from the solutions. 
In the abstract graph representation, a multi-particle emitter can be described by a hyperedge (shown in Fig. \ref{fig:448hyper} (c)). A hyperedge is drawn as a shape enclosing the $n>2$ vertices it connects. Thus, a \textsc{Halo}-subgraph constructed from regular edges can also be represented by hyperedges. Hypergraphs describing quantum experiments involving genuine multiparticle emitters have been explored in \cite{hypergraphs} -- these consequences can now be explored in the context of \halo's effective multiparticle emitters.

\textbf{Application of Concept} -- At this point, we wondered about the broadness of \textsc{Halo}s application. Is it of singular use for the design of 4-dimensional GHZ states, or can it be employed in more scenarios?

\begin{figure}[t]
    \centering
    \includegraphics[width=0.35\textwidth]{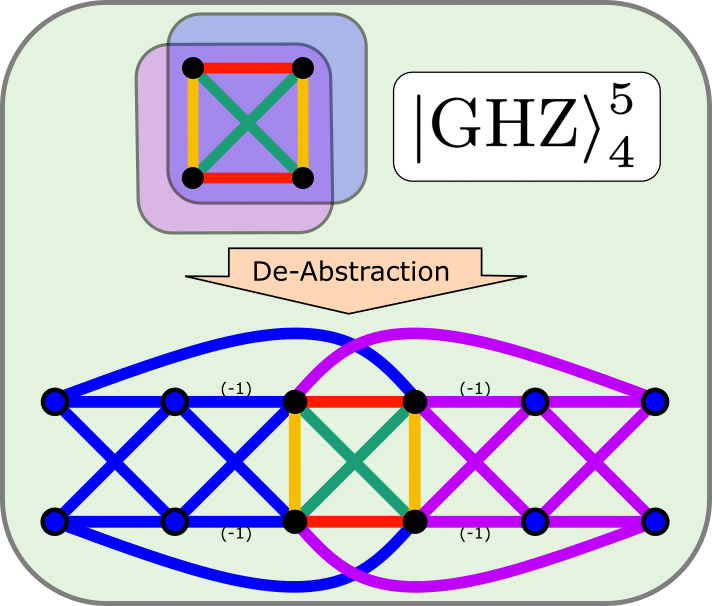}
    \caption{Example of how the graph for the creation of a $\ket{\text{GHZ}}_4^5$ state can be translated from the \textsc{Halo} picture. Since we know that a 4-hyperedge corresponds to a particular subgraph, the second hyperedge is just a recolored version of the first.}
    \label{fig:backtranslation}
\end{figure}

The definition made in the previous paragraph lets us extend the solution we have found. To show the definitions usefulness, we apply it to constructing generalizations of setups for high-dimensional (A) state creation, (B) entanglement swapping, and (C) quantum gates.\\
All three cases involve the following key steps:
\begin{enumerate}
    \item \textit{Discovery}: Discovering a graph for a quantum task with \pytheus.
    \item \textit{Abstraction}: Conceptualizing the underlying \textsc{Halo} structure.
    \item \textit{Utilization}: Using the \textsc{Halo} as a building block to construct generalized solutions by hand (no computation necessary).
\end{enumerate}
The last two steps show that we can constructively use the new concept without further computation, which is a crucial element for scientific understanding \cite{de2005contextual, potochnik2017idealization, deRegt2017}. \footnote{Formally, in de Regt's theory, scientific understanding is defined as follows\cite{deRegt2017}: A phenomenon P can be understood if there exists an intelligible theory T of P such that scientists can recognize qualitatively characteristic consequences of T without performing exact calculations.} 
\textbf{\textsc{H\footnotesize{ALO}} for Quantum State Creation} -- The \textsc{Halo}-subgraph shown in the previous paragraph creates an additional term to a four-particle state by imitating a four-particle emitter. This can be iterated with multiple copies of the same subgraph to produce states $\ket{\text{GHZ}}_4^d$ of arbitrary dimension $d$. The hyperedge representation is shown in Fig. \ref{fig:448hyper} (d).
The hypergraph constructed in this way can be deterministically translated back to reveal the underlying graph describing the corresponding experiment (see Fig. \ref{fig:backtranslation}). With this, we have found a general way of constructing experiments for the creation of high-dimensional multi-partite entanglement.

With \pytheus{}, we found another solution for a $\ket{\text{GHZ}}_6^3$ state with two ancillary particles that we can generalize to the state $\ket{\text{GHZ}}_{6+2n}^3$ for arbitrary $n\in \mathbb{N}$ using \textsc{Halo}. Another application is the creation of absolutely maximally entangled states, e.g. $\text{AME}(4,3)$.

\textbf{\textsc{H\footnotesize{ALO}} For Quantum Communication} -- Many quantum communication applications rely on the ability to distribute entanglement between different parties. Recently, new advanced have been made in chip-based quantum communication \cite{Wang2021-sr,chiptochipteleportation}. Here we apply \textsc{Halo} to general constructions of high-dimensional entanglement swapping experiments.

\begin{figure*}[t]
    \includegraphics[width=\textwidth]{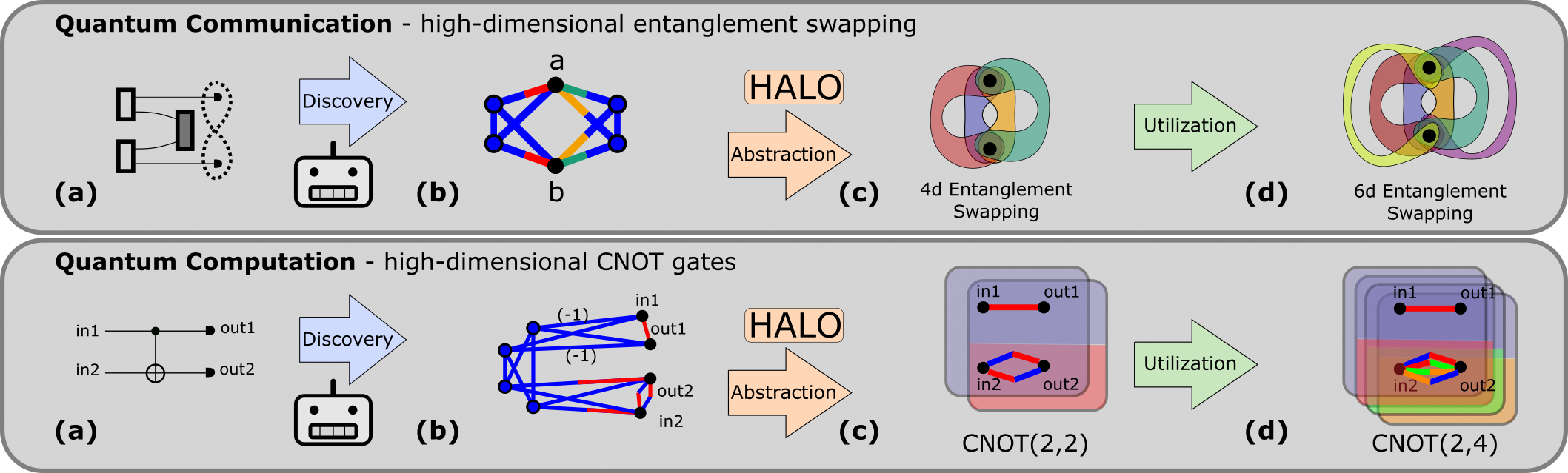}
    \caption{\textbf{Top} (a) shows a graph corresponding to two-dimensional entanglement swapping. (c) This graph has
two perfect matchings corresponding to two Bell terms (c) shows the solution for two-dimensional entanglement swapping in the abstract Halo representation. While the imitated (dashed) edges describe the creation of a Bell state, they do not correspond to any physical common source between the two particles reaching the detectors. (d) shows how the \textsc{Halo} interpretation can be used to build a four-dimensional entanglement swapping experiment using the same ancillary graph twice. \textbf{Bottom} (b) shows a graph corresponding to a photonic CNOT gate with four ancillas. (c) shows the solution in the abstract Halo representation. (d) shows how the \textsc{Halo} interpretation can be used to build photonic CNOT gates for arbitrarily high dimension.}
    \label{fig:commcomp}
\end{figure*}

In Fig. \ref{fig:commcomp} we show \pytheus's solution of an experimental setup for four-dimensional entanglement swapping between two particles. An entanglement swapping experiment corresponds to the task of creating a Bell state between particles \textbf{a} and \textbf{b} without a common source. In the graph representation this corresponds to the absence of connections between the vertices \textbf{a} and \textbf{b}. For details refer to \cite{hundred}. Much like the task of creating a $\ket{\text{GHZ}}_4^4$, this relies on ancillary particles. Instead of a four-particle emitter this ancillary subgraph emulates two two-particle emitters (shown in Fig. \ref{fig:commcomp} (c)).
Applying this \textsc{Halo} twice yields a graph for four-dimensional entanglement swapping using four ancillary particles. Applying this iteratively we can construct a $2k$-dimensional entanglement swapping experiment using $2k$ ancillary particles (where $k\in \mathbb{N}$). This is illustrated in Fig. \ref{fig:commcomp} (d). Analogously to state creation, the hypergraphs can be translated back into a graph corresponding to an experimental setup.\\
Similar approaches could also be taken to find general constructions for multi-particle entanglement swapping experiments.

\textbf{\textsc{H\footnotesize{ALO}} for Quantum Computation} -- We now apply these insights further in the study of photonic quantum gates. Photonic CNOT gates have been realized previously with linear optics \cite{cnotexp,PhysRevLett.126.130501}. Furthermore, their computer-inspired design was the subject of previous study \cite{computercnot}.
Our goal is to use \pytheus to find a suitable \textsc{Halo} for a general construction of high-dimensional CNOT gates. 
The action of the most general high-dimensional CNOT on a two-qudit state can be written as
\begin{align*}
    \text{CNOT}(d_1,d_2) \ket{m,n} = \ket{m, n+m \mod d_2},\\
    \text{for } 0\leq m <d_1 \text{ and }0\leq n <d_2,
\end{align*}
where $d_1$, $d_2$ are the dimensions of the two particles and $m$, $n$ are their modes.
When describing a quantum gate, incoming photons are represented by vertices just like detectors. An edge between an input vertex and a detector vertex symbolizes an incoming photon traveling to a detector. Edges between incoming photons are not permitted. For details refer to \cite{hundred}.
In Fig. \ref{fig:commcomp} (b) we show a $\text{CNOT}(2,2)$ gate between two qubits found with the help of \pytheus. While it is possible to create a CNOT with fewer experimental resources (two ancillary particles instead of four) our goal was to find a solution that can be generalized by interpreting it as a \textsc{Halo}.
\\
The subgraph of edges connected to the ancillary vertices corresponds to a part of the physical setup that encodes the action of the gate if \textbf{in1} enters in mode zero.
This principle can be applied further by duplicating the structure of the ancillary graph to extend the gate to a qubit controlling a four-dimensional qudit. Iterating this procedure yields a general construction of $\text{CNOT}(2,2k)$ gates for a controlled particle with an arbitrarily high dimension $2k$. This construction requires $4k$ ancillary particles. While more efficient use of experimental resources is possible \cite{hundred}, this solution provides a clear design rule for a gate $\text{CNOT}(2,2k)$ that can be directly translated into an experimental setup. Similar approaches to find extendable designs could also be taken to produce $\text{CNOT}(3,D)$ gates or even more generally $\text{CNOT}(D_1,D_2)$ gates.

\textbf{Outlook} -- \label{sec:conclusion}
In this article, we have presented \halo{}, a digitally discovered concept that experimentallly emulates probabilistic multi-pair emitters using only pair sources. We have shown the physics \halo{}s and how they can be used for general constructions of quantum experiments. The interpretation of the state creation \halo as emulating a multi-particle emitter opens a range of possible applications. Further, \halo{}s for emulating six- and eight-photon emitters have been found and hint at an exciting zoo of structures for creating entanglement. We have demonstrated how results found by a computer and their abstraction can be fruitfully applied and increase human understanding in tasks where pure human intuition fails. The three steps presented in this article provide a framework for human-computer cooperation, which delivers designs for quantum experiments that go beyond what humans or computers could find individually. We consider the representation in the form of graphs a key factor in our successful generalization. It is both computationally efficient for the computer and interpretable by humans. An extension of our approach to other domains of physics will be an interesting future project. For example in the field of quantum circuit design, abstract representations in terms of graphs \cite{duncan2020graph} and graph-based information flow \cite{anand2022information} could allow for a direct translation of our approach. At a big picture, our approach can be seen as a source of inspiration for new ideas in quantum optics, and we show here that such artificial systems can lead to new, generalizable  scientific concepts and understanding.

\textbf{Acknowledgements} -- 
The authors thank Xuemei Gu, Leonhard M\"o{}ckl, and Jonas Landgraf for helpful discussions and comments on the manuscript.
\bibliography{biblio}
\end{document}